\documentclass[%
	prl,
	reprint,
	groupedaddress,
	frontmatterverbose,
	showpacs,preprintnumbers,
	nobibnotes,
	amsmath,amssymb,
	aps,
]{revtex4-1}

\usepackage{graphicx}
\usepackage[space]{grffile}
\usepackage{latexsym}
\usepackage{textcomp}
\usepackage{longtable}
\usepackage{tabulary}
\usepackage{booktabs,array,multirow}
\usepackage{amsfonts,amsmath,amssymb}
\usepackage{natbib}
\usepackage{etoolbox}
\usepackage{xcolor}
\usepackage{url}
\usepackage[citecolor=blue,citebordercolor=blue,linkbordercolor=green]{hyperref}
\usepackage[caption = false]{subfig}
\usepackage[utf8]{inputenc}
\usepackage[english]{babel}
\usepackage{physics}

\newif\iflatexml\latexmlfalse
\DeclareGraphicsExtensions{.pdf,.PDF,.png,.PNG,.jpg,.JPG,.jpeg,.JPEG}

\begin{document}

\title{Core Electrons in the Electronic Stopping of Heavy Ions}
\thanks{Citation: \href{https://doi.org/10.1103/PhysRevLett.121.116401}{Phys. Rev. Lett. \textbf{121}, 116401 (2018)}}
\author{Rafi~Ullah}
\email[Electronic address: ]{ullah1@llnl.gov}
\affiliation{CIC nanoGUNE,  Ave. Tolosa 76, 20018 Donostia-San Sebasti\'{a}n, Spain}
\affiliation{Departamento de F\'isica de Materiales, UPV/EHU, 
Paseo Manuel de Lardizabal 3, 20018 Donostia-San Sebasti\'an, Spain}

\author{Emilio~Artacho}
\affiliation{Theory of Condensed Matter, Cavendish Laboratory, University of Cambridge, Cambridge CB3 0HE, United Kingdom}
\affiliation{CIC nanoGUNE and DIPC, Ave. Tolosa 76, 20018 Donostia-San Sebasti\'{a}n, Spain}
\affiliation{Basque Foundation for Science Ikerbasque, Bilbao, Spain}

\author{Alfredo~A.~Correa}
\email[Electronic address: ]{correaa@llnl.gov }
\affiliation{Quantum Simulations Group, Lawrence Livermore National Laboratory, 7000 East Avenue, Livermore, California 94550, USA}

\date{\today}
\begin{abstract}
Electronic stopping power in the \(\mathrm{keV/\AA}\) range is accurately calculated from first principles for high atomic-number projectiles and the effect of core states is carefully assessed. The energy loss to electrons in self-irradiated nickel is studied using real-time time-dependent density functional theory. Different core states are explicitly included in the simulations to understand their involvement in the dissipation mechanism. The core electrons of the projectile are found to open additional dissipation channels as the projectile velocity increases. Almost all of the energy loss is accounted for, even for high projectile velocities, when core electrons as deep as \(2s^2 2p^6\) are explicitly treated. In addition to their expected excitation at high velocities, a flapping dynamical response of the projectile core electrons is observed at intermediate velocities. The empirical reference data are well reproduced in the projectile velocity range of \(1.0 - 12.0~\mathrm{atomic~units}\) (\(1.5-210~\mathrm{MeV}\)).
\end{abstract}

\maketitle

The dissipative processes in ion irradiation of matter are of primary interest from the fundamental physics point of view (a paradigmatic example of strongly non-equilibrium but quasistationary processes) as well as for technological applications (aerospace electronics \cite{Bagatin_2015}, future energy application materials \cite{Granberg_2016}, radiation based cancer therapies \cite{Levin_2005}, and material science \cite{Townsend_1987}). The most dominant channel of energy dissipation for a swift ion shooting through matter is to the electronic degrees of freedom of the target. The energy loss to the host electrons is formally known as electronic stopping power (\({\cal S}_e\)) and defined as the energy lost by the projectile per unit path length (\({\cal S}_e = -\frac{dE}{dx}\)). The electronic stopping power of light ions shooting through simple metals in the low-velocity regime [\(v < 1\) atomic units (a.u. hereafter)] has been relatively well understood within linear response \cite{Fermi_1947,Lindhard_1954,Ritchie_1959} and non-linear response formalisms \cite{Almbladh_1976,Echenique_1981,Echenique_1986,Echenique_1990}. The linear and non-linear response approaches have been refined  \cite{Campillo_1998,Pitarke_2000,Juaristi_2000,Nagy_2009,Winter_2003,
Gondre_2013,Koval_2013,Sigmund_2014} but essentially remained limited in their practical applicability to simple metals and simple ions \cite{Roth_03_2017,Roth_10_2017}. 

More sophisticated approaches such as the time-dependent tight-binding method \cite{Mason_2007,Race_2010,Race_2013}, linear-response time dependent density functional theory (LR-TDDFT) \cite{Shukri_2016},  and real-time (RT)-TDDFT  \cite{Pruneda_2007,Krasheninnikov_2007,Quijada_2007,Hatcher_2008,Correa_2012,Zeb_2012,
Ojanpera_2014,Ullah_2015,Schleife_2015,Wang_2015,Quashie_2016,Lim_2016,Reeves_2016,
Chang_2017,Yost_2017,Gang_2017} have been applied to the problem of electronic stopping power.  These approaches have been very successful in describing the electronic stopping power of systems exposed to light projectiles (H, He), with \({\cal S}_e\) values of the order of \(10~\mathrm{eV/\AA}\).
\par
In this work, we have studied the prototypical case of self-irradiated nickel (a Ni projectile shooting through a Ni target) using RT-TDDFT. The electronic stopping power of self-irradiated Ni is predicted to be in the range of \(\mathrm{keV/\AA}\) \cite{Ziegler_2010}. No material with \({\cal S}_e\) values in the \(\mathrm{keV/\AA}\) range, to the best of our knowledge, has ever before been simulated beyond linear response. Furthermore, the role of the core and semi-core electrons of the target even with light projectiles in the stopping process has been shown to be quite significant \cite{Shukri_2016,Zeb_2012,Schleife_2015,Yost_2017}. Yost \emph{et al}. \cite{Yost_2017}, while studying a H projectile in a Si target, have shown that an explicit treatment of semicore and core electrons of the target atoms is essential for the calculation of stopping power at high projectile velocities. Using all-electron calculations, they have highlighted the challenge of explicit treatment of core electrons within the pseudopotential approach. Ojanper\"{a} \emph{et al}. \cite{Ojanpera_2014} have shown that the core electrons of the projectile play an important role as well. In this study we investigate the full effect of core electrons for both the target and the projectile within RT-TDDFT.
\par  
Ni based alloys are known for their radiation tolerance \cite{Lu_2016}, thermal stability and optimal mechanical properties, making them promising candidate materials for next generation energy and aerospace applications \cite{Jin_2016, Zhang_2016,Levo_2017}.  
The presence of Ni in structural alloys is known to play an important role in mitigation of swelling under irradiation \cite{Bates_1981}. Nickel, along with iron and tungsten,  is the subject of extensive radiation damage research \cite{Osetsky_2015, Granberg_2016, Zhang_2017} for energy applications.
\par
\begin{figure}[ht]
\includegraphics[scale=0.95]{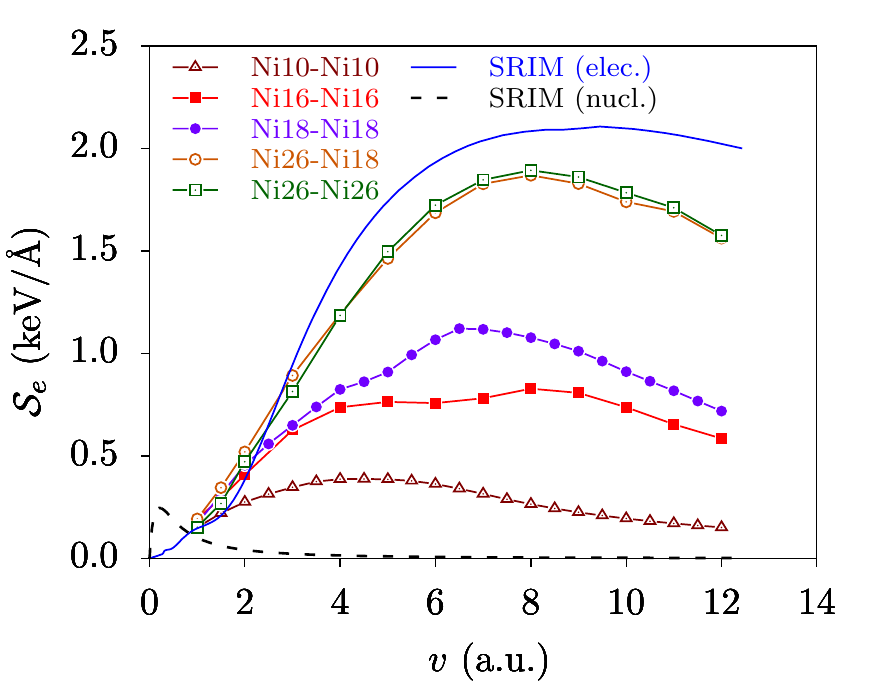}
\caption{Electronic stopping power (\({\cal S}_e\)) for a Ni projectile in a Ni crystal as a function of velocity.
For reference, both the nuclear (dashed, black curve) and the electronic  (solid, blue curve) stopping powers from SRIM \cite{Ziegler_2004} are presented. Open triangles (maroon) show calculated \({\cal S}_e\) for a Ni10 projectile in a Ni10 host,  solid squares (red) for a Ni16 projectile in a Ni16 host, filled circles (indigo)  for Ni18 in Ni18, open circles (orange) for Ni26 in Ni18, and open squares (green) for Ni26 in Ni26.} \label{fig:linearall}
\end{figure}

There are no direct experimental data available for the stopping power of Ni in Ni, except for the element-wise interpolations of Stopping and Range of Ions in Matter (SRIM) model \cite{Ziegler_2010}, which makes the prediction of our simulations ever more important. 
The SRIM model shows that in self-irradiated Ni, \emph{nuclear stopping} is dominant for velocities up to \(1~\mathrm{a.u.}\) and quickly diminishes beyond it (dashed curve in Fig.~\ref{fig:linearall}). We have considered the velocity regime \(1.0\) to \(12.0~\mathrm{a.u.}\) (\(1.5-210~\mathrm{MeV}\))  in which \({\cal S}_e\) becomes dominant and accounts for almost all of the total stopping power. 
\begin{table}[t]
	\begin{tabular}{ l | c  }
		Electronic configuration                                                                                  & Pseudopotential label           \\ [1ex]\hline   
		\fcolorbox{black}{black!20}{$1s^2 2s^2 2p^6 3s^2 3p^6$} \fcolorbox{black}{white}{$4s^2 3d^8$}             & Ni10   \rule{0pt}{5ex} \\ [2ex]
		\fcolorbox{black}{black!20}{$1s^2 2s^2 2p^6 3s^2$} \fcolorbox{black}{white}{$3p^6 4s^2 3d^8$}             & Ni16                   \\ [2ex]
		\fcolorbox{black}{black!20}{$1s^2 2s^2 2p^6$} \fcolorbox{black}{white}{$3s^2 3p^6 4s^2 3d^8$}             & Ni18                   \\ [2ex]
		\fcolorbox{black}{black!20}{\strut$1s^2$} \fcolorbox{black}{white}{\strut$2s^2 2p^6 3s^2 3p^6 4s^2 3d^8$} & Ni26                   \\ [2ex]
		\multicolumn{1}{c}{	\fcolorbox{black}{black!20}{ \strut core } \, \fcolorbox{black}{white}{\strut valence} }  &                        \\
	\end{tabular}
	\caption{Different pseudopotentials and labels utilized in this work. The number next to the element name indicates the number of \emph{explicit} electrons per atom.}
	\label{tbl:pseudo}
\end{table}
\par
 TDDFT is a reformulation of the many-electron time-dependent Schr\"odinger equation \cite{Runge_1984} analogous to what DFT is to the time-independent Schr\"odinger equation \cite{Hohenberg_1964}. Using the Kohn-Sham scheme, the many-body time-dependent problem is effectively reduced to a one-body problem in an effective potential \cite{Kohn_1965}. TDDFT is, in principle, exact; but in practice the exchange and correlation part of the effective potential is approximated using different schemes. 
In RT-TDDFT, the one-body Kohn-Sham wavefunctions are explicitly propagated in time, unlike what happens in the linear response approaches, which work in the frequency domain. We have used the RT-TDDFT formalism using the 
first-principles code \textsc{qb@ll} \cite{Gygi_2008,Draeger_2017} for our calculations, as described in Ref. \cite{Schleife_2012}.   The exchange and correlation within the adiabatic limit are obtained using the local density approximation \cite{Ceperley_1980}.
 It is known that dynamic exchange and correlation effects do play a role in the electronic stopping of ions in jellium in the low velocity limit \cite{Nazarov_2005,Nazarov_2007,Nazarov_2008}. Although these effects should be further investigated, the scale of the known corrections
as described in Refs. \cite{Nazarov_2005,Nazarov_2007,Nazarov_2008} is negligibly small for the velocity and stopping regimes
considered in this work.
\par
The Kohn-Sham wavefunctions represent individual electrons and are expanded in a plane-wave basis. \({\cal S}_e\) changes less than 3\% in the worst case as the energy cutoff is varied from \(160\) to \(400~\mathrm{Ry}\) (Supplemental Material \cite{supplemental}). Hence, all the calculations are performed using an energy cutoff of \(160~\mathrm{Ry}\). The ions are represented by norm-conserving non-local pseudopotentials, factorised in the Kleinman-Bylander form \cite{Kleinman_1982}. 
A supercell containing \(108\)~atoms was constructed by 
\(3\times 3\times 3\) conventional cubic cells of Ni. The size effects are discussed in Refs. \cite{Schleife_2015,Correa_2018}.
The experimental value of \(3.52~\mathrm{\AA}\) for the lattice constant was used. 
\par
The simulations could be described as virtual experiments (see Supplemental Material \cite{supplemental} for an actual video of the simulation). 
A Ni interstitial is placed inside the supercell and a self-consistent ground state is obtained. 
The self-consistent ground state serves as an initial state for the real-time evolution of the Kohn-Sham wavefunctions. 
From the self-consistent ground state, the Ni interstitial is instantaneously given a velocity at \(t=0\) mimicking an incident particle. As the projectile shoots through the bulk, the Kohn-Sham wavefunctions are propagated in time using a fourth order Runge-Kutta integrator \cite{Schleife_2012}, with all atoms fixed except the projectile, which moves with a constant velocity. The constrained ionic motion is based on the fact that ionic velocities, for the considered simulation times and trajectories, do not change significantly (\(\le1.0^{-4}~\mathrm{a.u.}\)).
After testing the convergence of simulation parameters, a time step of \(0.2~\text{attoseconds}\) or smaller is used for time-integration (\(\Delta t = \frac{\Delta x}{v}\) by additionally requiring \(\Delta x \leq 0.01~\mathrm{a_0}\)). The sudden kick causes a relatively short-lived transient before the system enters a dynamical steady state. The total Kohn-Sham energy of the electronic sub-system is recorded as a function of distance travelled by the projectile for different velocities (see Ref. \cite{Schleife_2015} for a discussion on the definition of the energy in the context of time-dependent Kohn-Sham equations).
The constrained motion of ions guarantees that the change in the electronic total energy along the trajectory corresponds to the `electron-only' stopping (\({\cal S}_e\)) experienced by the projectile. 
The slope of each of those curves is obtained by simple linear curve fitting as detailed in Refs.~\cite{Ullah_2015,Schleife_2015,Quashie_2016}, which gives \({\cal S}_e\) for that particular velocity. Equivalently, \({\cal S}_e\) can be calculated from forces acting on the projectile. Although it is a more general approach and gives the same stopping power values, it is computationally more expensive as forces converge at a smaller time step. The calculations in this work are in channeling conditions along the [111] direction of the face-centered cubic crystal of Ni.  To underline the importance of computational resources needed for similar calculations, it is worth mentioning that the computation of a single stopping power value for a given velocity takes about 200 000 CPU hours at an IBM BG/Q supercomputer \cite{Haring_2012}.

\par
We have investigated the contribution of core states by controlling their inclusion via a sequence of different pseudopotential approximations. 
This approximation replaces core electrons by an effective potential that reproduces the physics of the valence electrons. It is, in general, a necessary approximation when working with a plane wave basis \cite{Kohanoff_2006}. 
The core states frozen into a pseudopotential cannot polarize or take part in any dynamical process. Redefining the partition between valence and core electrons allows us to assess the approximation.
We have exploited this freedom to study the participation of the different core states in the process of energy deposition. 
We have generated four pseudopotentials, namely, Ni10, Ni16, Ni18, and Ni26 with different valence electrons, as defined in Table~\ref{tbl:pseudo} \cite{Fuchs_1999,Rappe_1990,Ilya_2000}.

\begin{figure}[ht]
\includegraphics[scale=0.90]{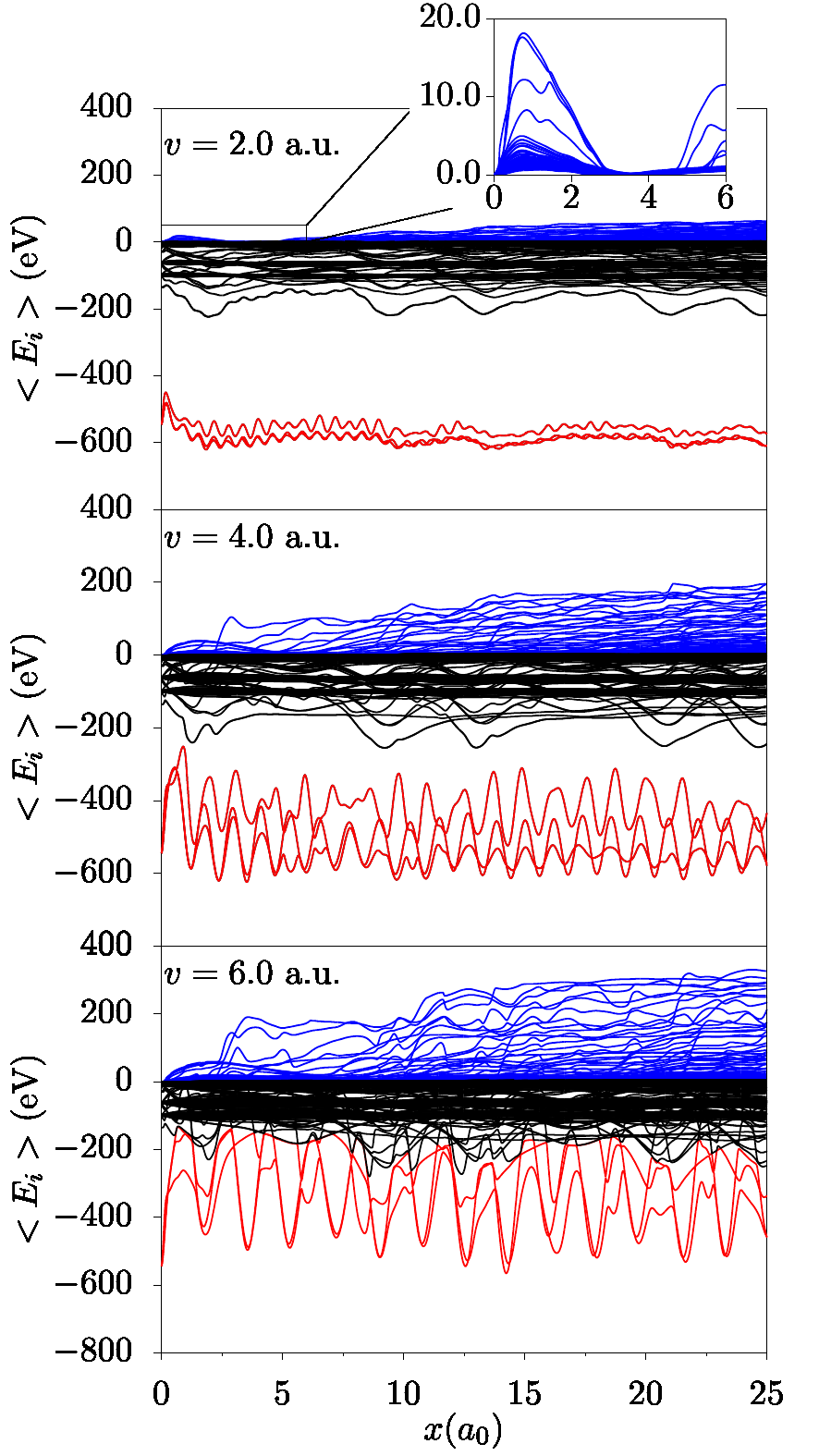}
\caption{
Instantaneous energy expectation values for the propagating Kohn-Sham states \( \expval{H_\text{KS}}{\psi_n^\text{KS}(t)}\) as a function of projectile position (relative to the unperturbed Fermi energy).
The red curves show \(2p^6\) states of the projectile. The black curves show the rest of the projectile states and the host states below the Fermi energy. The blue curves indicate as the states cross above the Fermi energy.
The inset in the top panel shows the length scale of the initial transient, due to the initial velocity kick.}
\label{fig:expect}
\end{figure}
\par
The results of our calculations for the different core or valence sets are presented and compared with the SRIM data in Fig.~\ref{fig:linearall}.
The calculated \({\cal S}_e\) of Ni10 in Ni10 
(Ni projectile and host atoms all with 10 explicit electrons)
is clearly underestimated in practically the whole velocity range investigated, (open triangles), by about an order of magnitude as compared to SRIM data (solid line). 
Not only is the \({\cal S}_e\) underestimated, but the maximum of \({\cal S}_e\) occurs around \(5~\mathrm{a.u.}\)
of velocity while SRIM predicts it to peak around \(9~\mathrm{a.u.}\). However, redefining more electrons from the frozen core to explicitly simulated valence states makes a very significant difference. 
In a similar calculation with a Ni16 projectile in a Ni16 host, the calculated \({\cal S}_e\) increases almost by a factor of two, as shown by the solid squares. This is a strong direct evidence ratifying the expected importance of core states
in the energy dissipation mechanism.
However, the \({\cal S}_e\) remains underestimated in comparison to the SRIM data.
Digging further in the same direction; we have calculated the \({\cal S}_e\) of Ni18 in Ni18 and Ni26 in Ni26. 
The Ni18 projectile in a Ni18 host calculation (solid circles), confirms the trend, although it does not fully account for the underestimation in the \({\cal S}_e\). 
The Ni26 projectile in a Ni26 host case (open squares) produces the \({\cal S}_e\), in perfect agreement with the SRIM data from \(1.0\) to \(3.0~\mathrm{a.u.}\) of velocity, while it is underestimated by less than 10\% between \(3.0\) to \(9~\mathrm{a.u.}\), which is within the anticipated inaccuracy in the SRIM model for heavier elements \cite{Ziegler_2004}.
\par
In addition to the good agreement with the SRIM model based data, these results provide a very clear evidence that core states as deep as \(2s^2 2p^6\) very significantly affect the \({\cal S}_e\) of the swift ions. The \({\cal S}_e\) values for different valence states converge in the low-velocity limit, but those for limited valence states saturate too early. The smaller the number of valence electrons, the earlier the \({\cal S}_e\) saturates with increasing velocity. Very importantly,
Fig.~\ref{fig:linearall} also reveals that if the right number of core electrons are allowed to participate in the dynamic processes, almost all of the dissipation can be accounted for within the RT-TDDFT formalism. 
\par
To distinguish the effect of core electrons in the
host from those of the projectile, we have computed the \({\cal S}_e\) of a Ni26 projectile in a Ni18 host (open squares). It is very interesting to note that it almost exactly matches the \({\cal S}_e\) of the Ni26 in the Ni26 case. 
The only difference between Ni18 in Ni18 (solid circles) and Ni26 in Ni18 is the presence and consideration of \(2s^2 2p^6\) as dynamical electrons of the projectile, which increases the \({\cal S}_e\) by a factor of almost two pointing to the importance of bare charge of the highly ionized projectile. This result strongly suggests that the critical contribution comes from the \(2s^2 2p^6\) electrons \emph{of the projectile} while the deep electrons of the host do not make a significant difference.
\par
Regarding the position of the peak of the \({\cal S}_e(v)\) curve, as more core electrons are treated explicitly, the peak position gradually corrects by shifting rightwards. 
The SRIM data predict the \({\cal S}_e\) peak position around \(9.4~\mathrm{a.u.}\) 
of velocity,  while our calculations put it around \(8.0~\mathrm{a.u.}\) of velocity, a 15\% smaller value.
\begin{figure}[ht]
\centering
\includegraphics[scale=0.95]{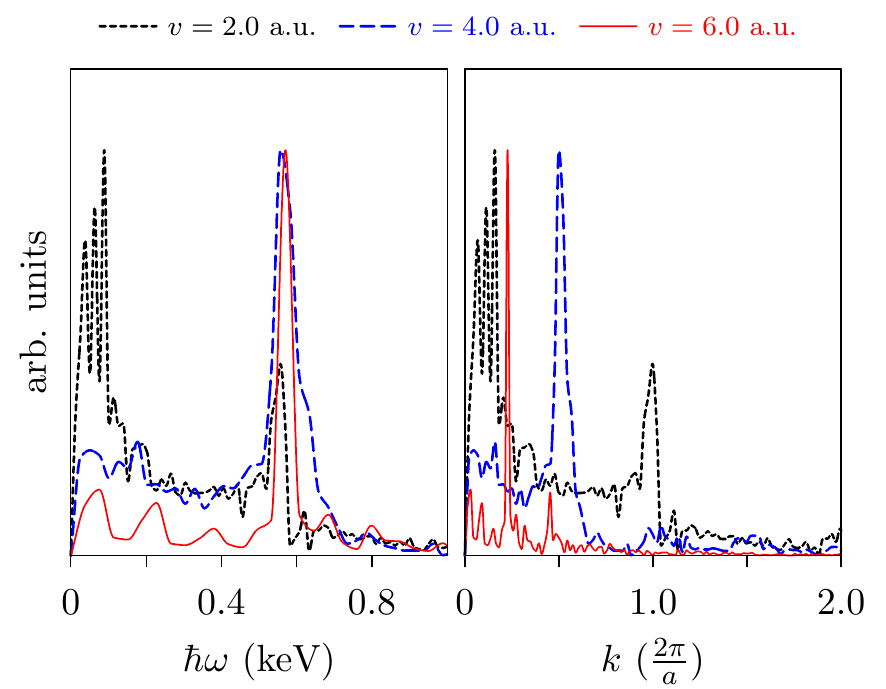}
\caption{Fourier transform of the energy of the 2\(p\) states (red curves of Fig. \ref{fig:expect}) in time (left panel, frequency expressed in energy units) and space (right panel).}
\label{fig:fourier}
\end{figure}

\begin{figure}
\centering
\subfloat{\raisebox{0.4in}{\rotatebox[origin=b]{270}{\(v = 2.0~\mathrm{a.u.}\)}}}
\hspace{1ex}
\subfloat{\includegraphics[scale = 0.07]{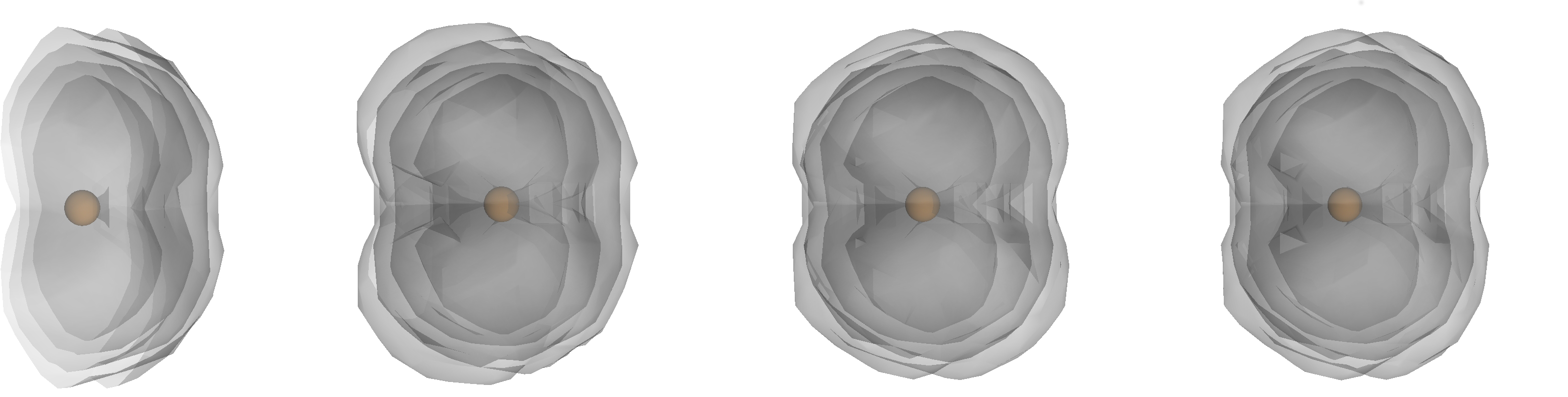}}\\
\subfloat{\raisebox{0.4in}{\rotatebox[origin=b]{270}{\(v = 4.0~\mathrm{a.u.}\)}}}
\hspace{1ex}
\subfloat{\includegraphics[scale = 0.07]{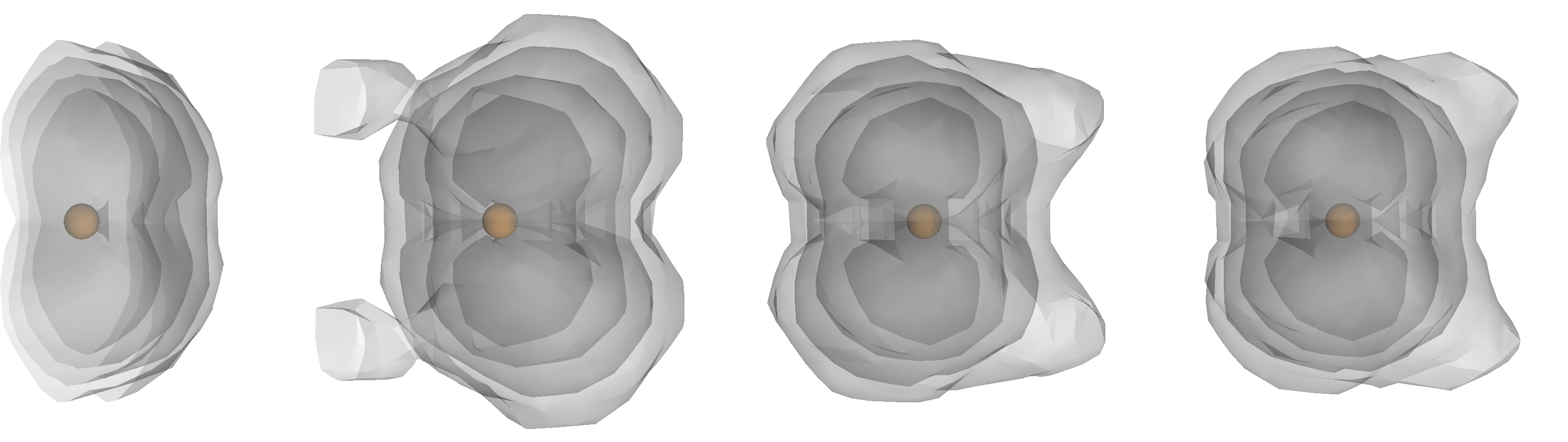}}\\
\subfloat{\raisebox{0.4in}{\rotatebox[origin=t]{270}{\(v = 6.0~\mathrm{a.u.}\)}}}
\hspace{1ex}
\subfloat{\includegraphics[scale = 0.07]{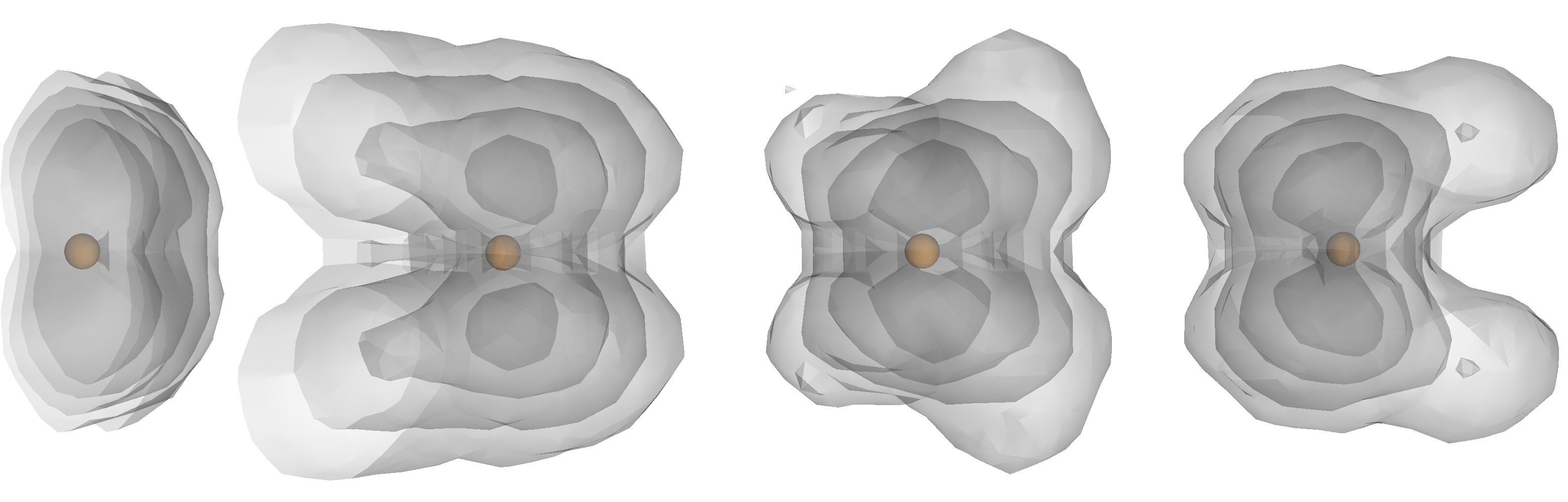}}\\
\subfloat{\raisebox{0.1in}{\rotatebox[origin=b]{0}{\(0.0~a_0\)}}}
\hspace{1cm}
\subfloat{\raisebox{0.1in}{\rotatebox[origin=b]{0}{\(0.80~a_0\)}}}
\hspace{1cm}
\subfloat{\raisebox{0.1in}{\rotatebox[origin=b]{0}{\(1.60~a_0\)}}}
\hspace{1cm}
\subfloat{\raisebox{0.1in}{\rotatebox[origin=b]{0}{\(2.40~a_0\)}}}
\caption{ The contour plot of \(2p\) orbital of the projectile. The orbital at the initial position appears clipped because the projectile is initially placed at (011) plane of the supercell. The orbital is plotted for three different projectile velocities starting from the same initial position with three subsequent projectile positions as it travels through the Ni crystal.}
\label{fig:2pallv}
\end{figure}
\par
The case of Ni26 projectile in a Ni18 host allows us to characterize the dynamics of the core electrons of the projectile. In Fig. \ref{fig:expect} we show the time evolution of the energy expectation values of the TDKS Hamiltonian for the occupied TDKS orbitals for different projectile velocities. The lowest levels can be identified as \(2s\) and \(2p\) states of the projectile (Ni26) in a Ni18 host, however \(2s\) is out of scale and not shown in the figure. The three lowest states in the figure (red curves) are \(2p\) states of the projectile. Although the calculation of \({\cal S}_e\) is well converged with respect to the energy cutoff, the quantitative convergence of individual core states would require higher cutoff energies; they do offer a good qualitative insight, however. Two distinct features, depending on the velocity regime, are immediately noticeable.
At low velocities the energy expectation values of core occupied levels of the projectile (red curves) do not change significantly, while the valence band shows that some dynamical states (blue curves) acquire energies that eventually reach hundreds of eV above the Fermi energy, forming an increasing set of ballistic electrons. These electrons would be ejected from the sample if they reach the surface
(the work function of Ni is \(\sim 5\) eV).
 At high velocity the latter effect is more pronounced, both in the number of electrons and the energy scale. More importantly, we see an effect that is absent at low velocity, related to the excitation of core electrons of the projectile into valence band energies and further into the ballistic range.
\par
Oscillations are observed in the energies of the projectile's 2$p$ core-state in Fig. \ref{fig:expect} at low to intermediate velocities. It is important to note that
these oscillations are not related to the lattice periodicity, but change with velocity as shown in the right panel of Fig. \ref{fig:fourier}, rather maintaining a constant period in time as shown in the left panel of Fig. \ref{fig:fourier}. This indicates that the oscillations are intrinsic to the dynamical process rather than to the external (crystal) periodicity. This behavior could be seen as a flapping of the core electrons as shown in Fig. \ref{fig:expect}, with a dynamical re-shaping in real space illustrated in Fig. \ref{fig:2pallv}. To quantify the dynamical re-shaping, we define \(\Delta(t) = x_\text{p}(t) - \langle \psi_{2p}(t) | \hat x | \psi_{2p}(t)\rangle\) where \(x_\text{p}(t)\) is the projectile position. \(\Delta(t)\) oscillates within \(\pm0.1~a_0\) for  \(v = 2~\mathrm{a.u.}\), and within \(\pm0.2~a_0\) for \(v = 6~\mathrm{a.u.}\)
\par
In summary, for Ni, like other transition metals that show a very high electronic stopping power, core electrons were found to have a major contribution in it, particularly those of the projectile. Adding explicit electrons in the simulation has the dual effect of adding more excitation channels, mainly in the form of electrons of the host, and making the ion potential deeper when ionization occurs, mainly in the projectile. While considering only 10 dynamical electrons per atom with frozen core could be a good approximation below \(1~\mathrm{a.u.}\) of velocity, the 18 electron approximation is valid up to \(2~\mathrm{a.u.}\), before saturating. For larger velocities, more electrons need to be taken into account to reproduce a reasonable value for the stopping power;
especially for the projectile ion including its core electron flapping behavior.

We are thankful to T. Ogitsu for making the Ni26 pseudopotential available. R. U. and E. A. would like to acknowledge financial support from MINECO-Spain through Plan Nacional Grants No.\ FIS2012-37549 and FIS2015-64886, and FPI Ph.D. Fellowship Grant No.\ BES-2013-063728, along with the EU Grant ``ElectronStopping" in the Marie Curie CIG Program. Work by R. U. (during a visit hosted by A. A. C.) and by A. A. C. was performed under the auspices of the U.S.
Department of Energy by Lawrence Livermore National
Laboratory under Contract No. DE-AC52-07NA27344
with computing support from the Lawrence Livermore
National Laboratory Computing Grand Challenge program,
and supported as part of the Energy Dissipation to Defect
Evolution (EDDE), an Energy Frontier Research Center
funded by the U.S. Department of Energy, Office of
Science, Basic Energy Sciences.

\end{document}